\documentclass[aps,floats,twocolumn,epsf,prl,showpacs]{revtex4-1}
\usepackage{graphicx}
\usepackage{epstopdf}
\usepackage{amsmath}
\usepackage{amssymb}
\usepackage[colorlinks=true,urlcolor=blue,citecolor=blue,linkcolor=blue,breaklinks=true]{hyperref}
\usepackage{color}
\usepackage[dvipsnames]{xcolor}
\usepackage[normalem]{ulem}

\begin{document}

\title{Two distinct quantum critical behaviors \\ in the doped two-dimensional periodic Anderson model}

\author{M. Kitatani$^{a}$, T. Sch\"afer$^{b}$, A. A. Katanin$^{c,d}$, A. Toschi$^e$ and K. Held$^e$}

\affiliation{{$^a$}Department of Material Science, University of Hyogo, Ako, Hyogo 678-1297, Japan}
\affiliation{$^b$Max Planck Institute for Solid State Research, Heisenbergstra{\ss}e 1, 70569 Stuttgart, Germany}
\affiliation{$^c$Center for Photonics and 2D Materials, Moscow Institute of Physics and Technology, Institutsky lane 9, Dolgoprudny, 141700, Moscow region, Russia}
\affiliation{$^d$ Institute of Metal Physics, 620990, Kovalevskaya str. 18, Ekaterinburg, Russia}
\affiliation{$^e$Institute of Solid State Physics, TU Wien, 1040 Vienna, Austria}

\date{\today}

\begin{abstract}
We study quantum criticality in the doped two-dimensional periodic Anderson model with the hybridization acting as a tuning parameter. Employing the dynamical vertex approximation we find two distinct quantum critical behaviors. One is a quantum critical point between the antiferromagnetically ordered and the Kondo state, both metallic with itinerant $f$ electrons. Here, we obtained the critical exponent $\gamma \approx 1$ for the temperature dependence of the antiferromagnetic susceptibility. We observe a \emph{second} quantum critical behavior with  $\gamma=2$  above the continuing zero-temperature magnetic order, at a quantum critical point  where the $f$ electrons turn from localized to itinerant.
\end{abstract}

\maketitle


{\sl Introduction}---When a second-order phase transition is driven down to zero temperature,
temporal (quantum) fluctuations play an essential role in addition to the thermal (classical) spatial fluctuations of the finite temperature phase transition. This  leads to 
a new class of universality at the zero-temperature quantum critical point (QCP) and 
a funnel-shaped  quantum critical region at finite temperatures above it.
In proximity of this quantum critical region exotic phenomena such as non-Fermi liquid behavior and  unconventional superconductivity may emerge \cite{Sachdev1999,coleman2005quantum,Loehneysen2007,Gegenwart2008a}.

Experimentally, the physics of QCPs has been mostly explored in heavy fermion systems 
\cite{Brando2016}, where the competition between the Ruderman-Kittel-Kasuya-Yosida (RKKY) magnetic interaction and the Kondo screening leads to a QCP.
The standard theory for such a magnetic QCP is the Hertz\cite{Hertz1976}-Millis\cite{Millis1993}-Moriya\cite{Moriya1973} (HMM) theory.
However, HMM theory is based on the effective bosonic model and  weak coupling, 
which is definitely not appropriate for heavy fermion
materials such as CeCu$_{\text{6-x}}$Au$_{\text{x}}$ \cite{Schroeder2000}, YbRh$_{\text{2}}$Si$_{\text{2}}$ \cite{Custers2003,Paschen2004} and Ce$_{\text{3}}$Pd$_{\text{20}}$Si$_{\text{6}}$ \cite{custers2012destruction,martelli2019sequential} that all have 
strongly correlated  $f$ electrons in common.

\begin{figure}[t]
        \centering
                \includegraphics[width=8cm,angle=0]{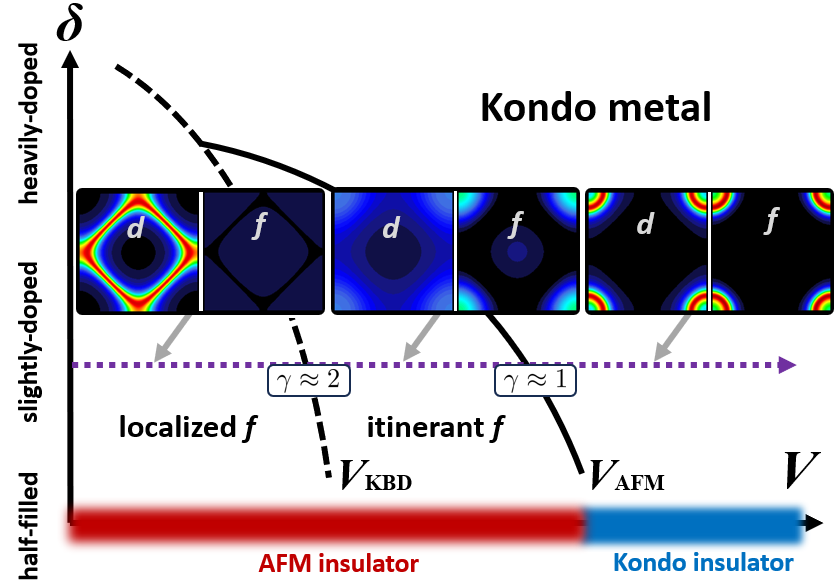}
        \caption{(Color online) Schematic phase diagram doping $\delta$ vs.  hybridization $V$ in the 2D  PAM for  heavy fermion systems.
        Three different phases, a paramagnetic Kondo metal and two AFM ordered phases with localized and with itinerant $f$ electrons are found; the inset shows their calculated Fermi surfaces  for $V=1.15t$, $V=0.9t$ and $V=0.7t$ (To enhance visibility, the $d$ electron spectrum is magnified by a factor of 10.).
        At half-filling, the system  turns from a metal to a Kondo and AFM insulator, respectively.
        \label{Fig:schematic}}
\end{figure}

The inadequateness of the HMM theory becomes most obvious at the
Kondo breakdown (KBD) transition, where the $f$ electrons change from itinerant to localized \cite{Si2003,Senthil2003,Coleman2005b,hoshino2013itinerant}.
This KBD is accompanied  by a dramatic change of the Fermi surface, which now needs to accommodate an additional (turned-itinerant) $f$ electron. In experiment, the change of the Fermi surface can also be measured by the Hall conductivity, which shows a jump \cite{Paschen2004}.

An intriguing question is whether the KBD QCP occurs 
together with the,
e.g., antiferromagnetic (AFM) QCP or is decoupled from it. Here,
all scenarios are possible: (i) the classical Doniach one with a single QCP with  simultaneous KBD and AFM  ordering \cite{Doniach1977,schaefer2019};
(ii) the occurrence of a KBD within the magnetic phase 
 \cite{hoshino2013itinerant,Raczkowski2022} as, e.g., in CeRh$_{1-x}$Co$_x$In$_5$ \cite{Goh2008}, CeRu$_2$(Si$_{1-x}$Ge$_x$)$_2$ 
 \cite{Matsumoto2011}, and CeRh$_{0.5}$Ir$_{0.5}$In$_5$ \cite{Kawasaki2020}; and (iii)
 a KBD within the paramagnetic phase preceding magnetic ordering
 as, e.g., in YbRh$_2$Si$_2$ \cite{Friedemann2009}. 
Fig.~\ref{Fig:schematic} depicts case (ii), as we found in the present paper  for the doped priodic Anderson model (PAM) and in \cite{Watanabe2007,Martin2008,Lanata2008}, with the subsequent occurrence of three different phases upon increasing $V$ at a slight doping:
 an AFM metal with localized $f$ electrons, an AFM metal with itinerant $f$ electrons and a Kondo metal.
  By using lattice frustration as an additional tuning knob beyond Fig.~\ref{Fig:schematic},
  (anti)ferromagnetism can be suppressed, and  one might be able to go from scenario (ii) to (i), and then to (iii) \cite{si2006global,si2014kondo}.

As for the quantum critical behavior,  which is at the focus of the present paper,
in case (i) where the AFM and KBD transition occur simultaneously, it has been discussed that the locally quantum critical state can change the critical exponent of the AFM QCP away from HMM theory \cite{si2001locally,Si2003}. 
In scenario (iii) an ideal local QCP without any interference by magnetism is expected and additionally a magnetic QCP   \cite{si2006global}.
If the  KBD occurs in the AFM region, instead, as in scenario (ii), it is expected that the quantum critical nature of the spin susceptibility by the Kondo destruction disappears, and only the classical AFM critical behavior is visible because the magnetic ordering occurs at a finite transition temperature and thus covers up the KBD. However, if the system is two-dimensional, the AFM transition temperature ($T_C$) should be zero even in the presence of an ordered ground-state because of the Mermin-Wagner theorem \cite{Mermin1966}. Then, the question arises: Does the KBD transition {\it within} the ordered groundstate give rise to some quantum critical feature at finite temperatures? Such a
QCP has, to the best of our knowledge, not been addressed before.

In this Letter, we study quantum criticality in  a representative model for heavy fermion metals: the periodic Anderson model (PAM)  at finite doping and in two dimensions (2D). We employ an advanced numerical approach, the dynamical vertex approximation (D$\Gamma$A) \cite{Toschi2007} with Moriya $\lambda$-correction \cite{Katanin2009,Rohringer2016}, which is a diagrammatic extension \cite{Rubtsov2008,Rohringer2013,Taranto2014,Ayral2015,Kitatani2015,Li2015,DelRe2019,RMPVertex} of dynamical mean field theory (DMFT) \cite{Metzner1989,Georges1992,Georges1996}. 
We calculate the quantum critical properties and find two distinct quantum critical behaviors for the scenario (ii) of Fig.~\ref{Fig:schematic}.

{\sl Method}---The D$\Gamma$A captures non-local spatial correlations beyond DMFT by taking into account particle-hole ladder diagrams  based on the DMFT irreducible vertex. 
One advantage of this scheme is that it can describe strongly correlated regimes, as a  major part of non-perturbative effects is already encoded in the local vertices of the self-consistent DMFT solution. From this local vertex the local DMFT correlations as well as non-local spin and charge fluctuations are constructed through the Bethe-Salpeter ladder diagrams in the particle-hole ($ph$) and transversal $ph$ channel. Particle-particle fluctuations are restricted to their local DMFT contributions. 

Among others, D$\Gamma$A describes the Mott and pseudogap physics of the two-dimensional Hubbard model \cite{Schaefer2015-2,Schaefer2021,Worm2024} and, if also the particle-particle channel is included, high-temperature superconductivity \cite{kitatani2019critical,kitatani2020nickelate}.
A particular advantage is that D$\Gamma$A  captures long-range spatial fluctuations, which are typically missed by cluster methods \cite{Maier2005} but expected to be essential when the system approaches a (quantum) critical point \cite{Sachdev1999,Katanin2009}. Let us recall that the D$\Gamma$A satisfies the Mermin-Wagner theorem \cite{Katanin2009,Rohringer2016,RMPVertex} for the antiferromagnetic susceptibility in 2D, which is crucial since the dimension is a key parameter for classifying the universality of QCPs.

For the above-mentioned  reasons D$\Gamma$A, while computationally expensive, is suitable for exploring quantum critical phenomena in low dimensional and strongly correlated electron systems. In fact,  D$\Gamma$A  and the closely related dual fermion (DF) approach have been successfully applied to classical and quantum critical behaviors in various correlated electron systems \cite{Rohringer2011,Schaefer2016,Antipov2014,Hirschmeier2015,DelRe2019}. For the two-dimensional particle-hole-symmetric  (and insulating) PAM, the D$\Gamma$A gives and scenario (i) and the critical exponent $\gamma \approx 2$, consistent with the field theoretical approach for two-dimensional (2D) quantum antiferromagnets, as well as the dramatically increase of the magnetic susceptibility in the crossover regime \cite{schaefer2019}.
This recent methodological progress now provides the  chance
for a better and unifying  understanding of QCPs in the PAM and in heavy fermion materials beyond the limits of hitherto existing theories.

{\sl Model}---We consider the 2D PAM at a finite doping. The ground state is metallic, and a completely different universality from the insulating case studied in \cite{schaefer2019} is expected. The Hamiltonian of the PAM reads 
\begin{align}
    {\cal H}_{\rm PAM} 
    =&
    \sum_{{\mathbf k}\sigma}
    (\epsilon_{\mathbf k}+\epsilon_{df}) 
    c^{\dag}_{{\mathbf k}\sigma}
    c^{\phantom{\dag}}_{{\mathbf k}\sigma}
    +U \sum_{i} n^{f}_{i \uparrow} n^{f}_{i \downarrow}
    \notag \\
    &+V\sum_{{\mathbf k}\sigma}
    (f^{\dag}_{{\mathbf k}\sigma} c^{\phantom{\dag}}_{{\mathbf k}\sigma} + h.c.) \; .
\end{align}
Here, $f^{\dag}_{i\sigma} (f_{i\sigma})$ is the creation (annihilation) operator for $f$ electrons with spin $\sigma \in \{ \uparrow, \downarrow \}$; $n^f_{i\sigma} \equiv f^{\dag}_{i\sigma}f^{\phantom{\dag}}_{i\sigma}$; $c^{\dag}_{i\sigma} (c_{i\sigma})$ is the creation (annihilation) operator for conduction electrons with dispersion $\epsilon_{\mathbf k} = -2t[{\rm cos}(k_x)+{\rm cos}(k_y)]$ and spin $\sigma$. Further, $V$ is the hybridization strength between conduction and $f$ electrons, $U$ the a onsite Coulomb repulsion, and $\epsilon_{df}$ is the energy difference between local conduction and $f$ electrons. 
In this study, we vary the hybridization $V$ at fixed $U=6t$ and  a slight doping $n_{\rm tot} \equiv n_d+n_f=1.9$, which is achieved by adjusting the chemical potential.  While the $f$ electron density $n_f$ is not an independent  tuning parameter, we set $\epsilon_{df}=3.4t$ to realize $n_f \approx 1$  so that we obtain the Kondo metallic state in the large $V$ region. Note that $n_{\rm tot}=2.0, \epsilon_{df}=3t$ would correspond the particle-hole symmetric half-filled and insulating PAM of  \cite{schaefer2019}. 

\begin{figure}[tb] 
        \centering
               \includegraphics[width=8.8cm,angle=0]{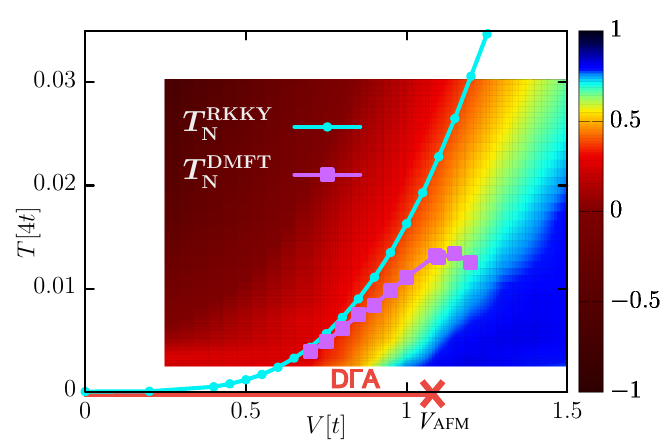}
        \caption{(Color online) Phase diagram temperature vs. hybridization strength in the 2D PAM at $U=6t$ and $n_{\rm tot}=1.9$.
          The figure shows the RKKY temperature $T_{\rm RKKY}$ (light-blue), the AFM N\'eel temperature $T_N$  in DMFT (purple). The red line marks the D$\Gamma$A  AFM order restricted to zero temperature, and the red cross the D${\Gamma}$A  QCP. The color code shows the exponent $\alpha$  extracted from the fit:
          $\Im \Sigma_{\rm DMFT}(\omega_n) \propto |\omega_n|^{\alpha}$ for low Matsubara frequencies.}
          \label{Fig:Doniach}
\end{figure}

\begin{figure*}[ht!]
        \centering
               \includegraphics[width=\textwidth,angle=0]{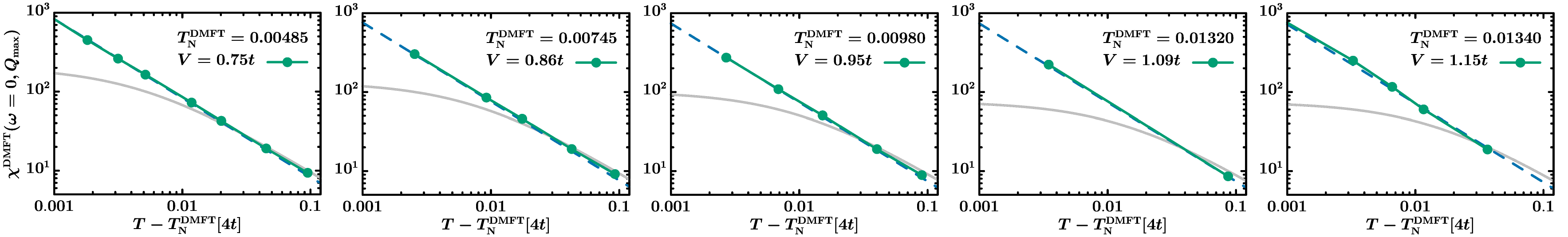}
                \vskip 0mm
                \includegraphics[width=\textwidth,angle=0]{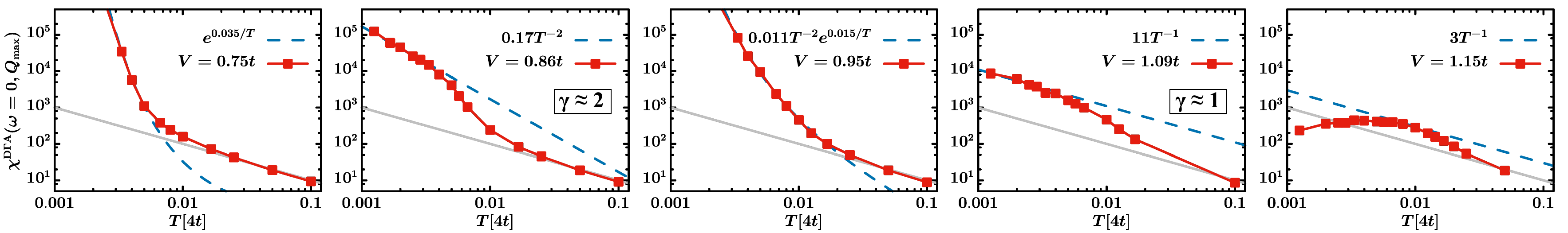}
        \caption{(Color online) Spin susceptibilities (on a log-log scale)  in DMFT (upper panel, green circles) and D$\Gamma$A (lower panel, red squares). The gray solid lines show the $\chi=T^{-1}$ Curie behavior. The dotted blue lines indicate the low temperature behavior of the susceptibility in DMFT [upper panel: $\propto (T-T_{\rm DMFT})^{-1}$] and D$\Gamma$A [lower panel], respectively.
        \label{Fig:chi} \label{Fig:susc}}
\end{figure*}


{\sl Phase diagram}---
Let us start by discussing the temperature $T$ vs.\ hybridization $V$ phase diagram which is shown in Fig.~\ref{Fig:Doniach}.
First, the light-blue line shows the RKKY antiferromagnetic transition temperature ($T_{\rm RKKY}$) of the mapped Kondo model with $J=2 V^2 [\epsilon_{\rm df}^{-1}+(U-\epsilon_{\rm df})^{-1} ]$ in the localized limit $(U \rightarrow \infty)$ \cite{Schrieffer1966}. Considering the second order perturbation-theory of J, we solve the following self-consistent equation to obtain $T_{\rm RKKY}$:
\begin{align}
    T_{\rm RKKY} = \frac{J^2}{4} \chi_{\rm d}^{\omega=0, \mathbf{Q}}(T_{\rm RKKY}),
    \label{Eq:TRKKY}
\end{align}
where $\chi^{\omega=0,\mathbf{Q}}_{\rm d}(T)$ is the $\omega=0$ susceptibility of conduction electrons at peak momentum $\mathbf{Q}$ with the density $n_d=0.9$ [we  assume $n_f=1$ for arriving at Eq.~(\ref{Eq:TRKKY})] and temperature $T$.

Second, the color map of  Fig.~\ref{Fig:Doniach} shows the extracted exponent $\alpha$ in the fit $\Im \Sigma^{\rm DMFT}(\omega_n) \propto |\omega_n|^{\alpha}$ to the two lowest Matsubara frequencies. For the normal Fermi liquid, $\Im \Sigma(\omega_n)$ behaves linearly (i.e., $\alpha = 1$). For localized $f$ electrons  $\Im \Sigma(\omega_n) \propto 1/\omega_n$ as in the atomic limit (i.e., $\alpha = -1$). We can thus see the crossover from the Fermi liquid (blue) region to the localized spin (red) region. The crossover temperature , roughly speaking the yellow region, 
is nothing but the Kondo temperature scale $T_K$ (in DMFT). 

Third, the purple line shows the DMFT N\'eel temperature. It matches $T_{\rm RKKY}$ for small $V$ where perturbation in $J\propto V^2$ is well controlled, and where RKKY dominates over the Kondo effect. However, since the
Kondo temperature  $T^{\rm DMFT}_{\rm K}$ grows exponentially with $V$, it eventually
surpasses $T^{\rm DMFT}_{\rm N}$.
The point $T^{\rm DMFT}_{\rm K} \approx T^{\rm DMFT}_{\rm N}$, i.e., the intersect of the yellow region with the purple line, marks the crossover from RKKY magnetic order at small $V$ to the Kondo screening at large $V$. At this point the $T^{\rm DMFT}_{\rm N}$ turns down, since the Kondo screening that couples the $f$ electron spin to the conduction electron spin competes with the 
$f$--$f$ spin coupling of the AFM RKKY phase.
Thus, Fig.~\ref{Fig:Doniach} essentially reproduces the well-known Doniach phase diagram \cite{Doniach1977}, which reflects that the DMFT well treats temporal fluctuations, including the Kondo physics (c.f.\ Ref.~\cite{Otsuki2009,Amaricci2012}).

Fourth, when we consider the non-local fluctuation effect in D$\Gamma$A, $T_N$ goes to zero due to the Mermin-Wagner theorem (red  D$\Gamma$A line). As a function of hybridization, we then find in Fig.~\ref{Fig:Doniach} a zero-temperature quantum phase transition at $V_{\rm AFM} \approx 1.09$ (red cross). This hybridization strength roughly corresponds to where the Kondo 
and the N\'eel temperature become comparable in DMFT or the 
and $T^{\rm DMFT}_{\rm N}$ starts turning down. 
It is slightly lower, however, since non-local correlations, that are included in  D$\Gamma$A, reduce the tendencies to  AFM order. These results for the doped PAM are  consistent with previous work for the insulating PAM at  half-filling \cite{schaefer2019}. 

{\sl Critical behavior of the spin susceptibility}---Next, we show the antiferromagnetic spin susceptibility $\chi^{\omega=0}_{\rm AF}$ at momentum ${\mathbf Q} \approx (\pi,\pi)$  
where the susceptibility takes a maximal value [In the low-temperature region, we obtained the crossover from commensurate $Q_{\rm max}=(\pi, \pi)$ for $V \gtrsim V_{\rm KBD}$ to incommensurate $Q_{\rm max}=(\pi,\pi-\delta)$ for $V \lesssim V_{\rm KBD}$. For $V<V_{\rm KBD}$, the incommensurability roughly reflects a nesting vector of the non-interacting d-bands with $n=0.90$]. 
We can see critical behaviors both in DMFT [Fig.~\ref{Fig:susc} (upper panels)] and D$\Gamma$A [Fig.~\ref{Fig:susc} (lower panels)]. In DMFT, we observed a crossover from Curie's law (i.e., free spin behavior: $\chi =T^{-1}$) at high temperature ($T \sim 0.1[4t]$) to  critical behavior 
 $\chi =(T-T_N^{\textrm{DMFT}})^{-\gamma} = (T-T_N^{\textrm{DMFT}})^{-1}$
at low temperature with a DMFT critical exponent $\gamma=1$.
This $\gamma$ reflects the (bosonic) mean-field critical behavior of DMFT, which neglects spatial fluctuations.

The D$\Gamma$A results  in Fig.~\ref{Fig:susc} (lower panels), while similar to DMFT for high temperature, display much richer properties at low $T$. First, we capture the change from the diverging  $\chi\sim e^{\Delta/T}$ behavior at  small $V (\lesssim 1.09t)$ 
to the Kondo screened behavior with a suppressed (screened) $\chi$ in the large $V$ region ($V\gtrsim 1.1t$).

In between, at the AFM QCP $V=V_{\rm AFM}\approx 1.09 t$, we find quantum critical behavior $\chi \propto T^{-1}$ whose exponent ($\gamma \approx 1$) agrees with HMM theory with dynamical exponent $z=2$ (for itinerant antiferromagnetic system) except for the logarithmic divergence term. 
At $V_{\rm AFM}$ and intermediate temperatures  ($T \sim 0.01[4t]$), we here observed the rapid enhancement of $\chi$ when crossing over from the high-$T$ 
$\chi = T^{-1}$ to the low-$T$ quantum critical behavior of susceptibility $\chi \approx11 T^{-1}$. 

Furthermore, within the antiferromagnetically ordered ground-state region, we observe a second power-law behavior $(\gamma \approx 2)$ at $V =V_{\rm KBD}\approx 0.86t$, which indicates a second  QCP. As will become clear later when looking at the Fermi surface, this is the KBD QCP.
The exponent $\gamma \approx 2$ matches the expected one for the insulating phase, for which (neglecting the presumably small anomalous exponent $\eta$) one should have  $z=1,\nu=1,\gamma=2$ \cite{PhysRevLett.60.1057,chubukov1994theory,schaefer2019}. 
Remarkably, the fitted (blue dotted) line at $V=0.86t$ indeed holds the relation $\chi^{\omega=0}_{\mathbf{Q}} \approx 20 J_{\rm H}/T^2$ where the magnetic coupling can be identified with the RKKY temperature, i.e., $J_{\rm H} \sim T_{\rm RKKY} = 0.009\;[4t]$, similar as for the insulating PAM (see Supplemental material of Ref.~\cite{schaefer2019}).
The fact that the critical behavior is well described by the mapped Heisenberg antiferromagnets suggests that the localized character of $f$ electrons dominates the KBD QCP within the AFM ordered phase, even though the KBD connects localized to itinerant $f$-electrons. It contrast to the $V_{\rm AFM}=1.09t$  QCP.

As it is multi-faceted, let us summarize the $V$ dependence of the low-temperature susceptibility $\chi_{\mathbf Q}(T)$ in D$\Gamma$A. For small $V<V_{\rm KBD}$, the susceptibility increases  exponentially with $1/T$  for $T \rightarrow 0$, consistent with the physics of 2D quantum antiferromagnets and of the 2D Hubbard model \cite{Chakravarty1988,Schaefer2015-2}. At $V=V_{\rm KBD}$, the susceptibility is reduced by the quantum fluctuations, and we observed $\chi \propto T^{-2}$, consistent with the quantum critical behavior for two-dimensional localized spins. Between $V_{\rm KBD}< V<V_{\rm AFM}$, 
the susceptibility shows an exponential increase with $1/T$ again  but now with an additional $T^{-2}$ prefactor, originating from itinerant f-electrons. We take this as a signature that the $f$ electrons change their behavior from localized to itinerant at $V=V_{\rm KBD}$.
Finally, at $V=V_{\rm AFM}$, we obtain itinerant antiferromagnetic critical behavior $\chi \propto T^{-1}$, and, eventually,  for $V>V_{\rm AFM}$ a Kondo screened susceptibility.

\begin{figure}[b]
        \centering
                \includegraphics[width=8.8cm,angle=0]{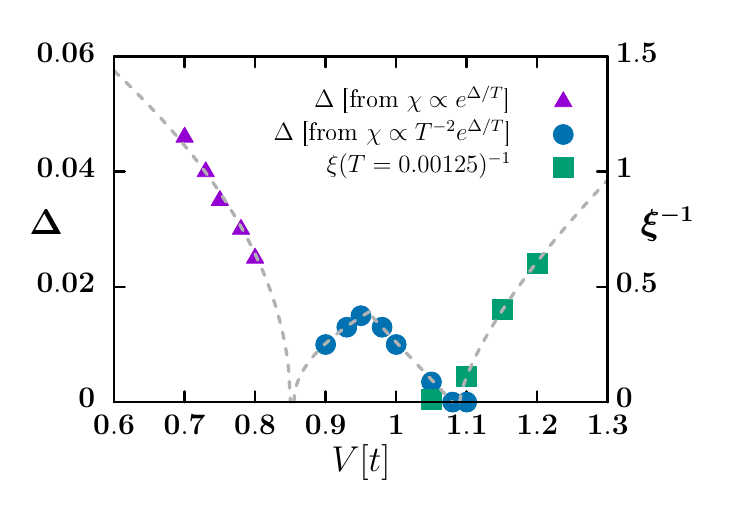}
        \caption{(Color online)  AFM energy scale $\Delta$ (fitted to the equations of the legend box) and correlation length $\xi$ as a function of  hybridization $V$. 
        \label{Fig:QC-phase}}
\end{figure}

{\sl Signature of two quantum critical regions}---To confirm two quantum critical behaviors, we plot  in Fig.~\ref{Fig:QC-phase} the energy scale of the AFM fluctuations $\Delta$  for $V \le V_{\rm AFM}$ (AFM phase) and the inverse correlation length $\xi^{-1}$ for $V \ge V_{\rm AFM}$ (PM phase).
Here, $\Delta$ is extracted from the fits $\chi \propto e^{\Delta/T}$ and $\chi \propto T^{-2}e^{\Delta/T}$ for $V \le V_{\rm KBD}$ and $V \ge V_{\rm KBD}$, respectively; $\xi$ is derived from $\chi(\mathbf{Q})^{-1} \propto (\mathbf{q}-\mathbf{Q})^2+\xi^{-2}$ (for details see the Supplemental Material \cite{SM}). Note that it is difficult to extract $\Delta$ near the QCP because the quantum critical behavior masks the exponential scaling. For this reason, we cannot reliably extract the exponent $\beta$ for $\Delta \propto |V-V_{\rm AFM(KBD)}|^{\beta}$.
Nonetheless, Fig.~\ref{Fig:QC-phase} clearly indicates two distinct QCPs: $V_{\rm KBD}\approx 0.86t$ and $V_{\rm AFM}\approx 1.09t$ \footnote{Since the calculations are performed at finite temperatures, we cannot fully rule out the possibility of a weak first-order transition, which would cause the quantum-critical-like behavior to end at some low temperature.}. At both QCPs the 
AFM energy scale $\Delta$ is suppressed.
At the second QCP from the  paramagnetic to the AFM  phase, located at $V_{\rm AFM}$,
we also observe a divergence of the correlations length $\xi$.

Further insight can be gained by looking 
at  the spectral weight at the Fermi level $A(\mathbf{k},\omega=0)$ calculated with the Pad\'{e} approximation which is shown for the three different regimes at the lowest temperature in our DMFT  in Fig.~\ref{Fig:schematic} as insets. (i) For small hybridization, $V=0.7t<V_{\rm KBD}=0.86t$, conduction and $f$ electrons are almost decoupled; $f$ electrons are akin to localized spins without spectral weight at the Fermi energy, and conduction electrons are $10\%$-hole doped from half-filling. That means we have a small Fermi surface of a bit less than half the Brillouin zone. Here, we are in the RKKY regime and the $f$ electron magnetism is almost equivalent to the Heisenberg antiferromagnet with an exponential scaling of the susceptibility $\chi \propto e^{\Delta/T}$.

(ii) For intermediate hybridization, $V_{\rm KBD}<V=0.9t<V_{\rm AFM}$, conduction and $f$ electrons start to mix visibly leading to a heavy fermion state. We have a single large Fermi surface which now must encompass 1.9 electrons in total, i.e., almost the entire Brillouin zone since the  $f$ electrons have become itinerant at $V_{\rm KBD}$.
The Fermi surface is  mixed (hybridized) however with 
a major $f$ contribution and a minor one from the  conduction electrons.  Here, also an exponential scaling of the susceptibility is found, but now with a $T^{-2}$ prefactor, i.e.,  $\chi \propto T^{-2}e^{\Delta/T}$. 

(iii) For large hybridization $V=1.15t>V_{\rm AFM}$, 
conduction and $f$ electrons are strongly coupled and form a joint Fermi surface (with $f$ electron still giving the main [$\sim 10$ times larger] contribution). Now, the localized $f$ electron's spin is screened in a Kondo singlet, and the susceptibility is thus suppressed for low temperatures.

 In between these three regimes, we obtained two quantum critical behaviors, whose exponents ($\gamma \approx 2$ and $\gamma \approx 1$) well corresponding to localized and itinerant $f$ electron characters, respectively. This behavior is summarized in Fig.~\ref{Fig:schematic}.
 
{\sl Conclusion}---We have examined the quantum critical behavior in the doped PAM, the textbook model for heavy fermion metals, by means of D$\Gamma$A. Thanks to the recent progress of numerical methods, we can now approach the quantum criticality of the arguably most difficult and debated case: two-dimensional strongly correlated itinerant antiferromagnetic metal with a Kondo breakdown.
There are two important mechanisms at work: (i) the competition between the RKKY interaction and the Kondo screening and (ii) the Kondo breakdown. We found two distinct quantum critical behaviors reflecting that these two mechanisms are decoupled. The first one at $V_{\rm AFM}$ is between the AFM phase and the paramagnetic Kondo metal. At this QCP, the exponent $\gamma \approx 1$ agrees with the conventional HMM theory for antiferromagnetic metals. The second quantum critical behavior at $V_{\rm KBD}$  with $\gamma \approx 2$ is due to the Kondo breakdown. Here, $\gamma \approx 2$ is consistent with quantum Heisenberg antiferromagnets reflecting the localized $f$ electron nature on one side of the QCP.  Such a QCP in the ordered regime in heavy fermion systems has only  been  discussed in the three-dimensional system before \cite{custers2012destruction,martelli2019sequential}. Because of the specialty of the two-dimensional system that magnetism is restricted to $T_N=0$, we are able to observe the magnetic quantum critical signature of the second QCP {\it inside} the region with an AFM  ground state. At the Kondo breakdown QCP, the antiferromagnetic energy scale $\Delta$ drops however towards zero (Fig.~\ref{Fig:QC-phase}).

Our result could serve as a  starting point for understanding the quantum critical phenomena of the global phase diagram of heavy fermion materials in an unbiased way. In the future, the calculations of real materials in three-dimensions and the investigation of the effect of weak three-dimensionality on our results are important subjects to study. Also whether the two QCP merge at a more heavy doping or when including lattice frustration, disorder or Fermi surface anomalies is a relevant question worthwhile studying.


{\sl Acknowledgments}---
The present work was supported in part by the
 Research Unit
QUAST by the
Deutsche Foschungsgemeinschaft (DFG project ID FOR5249), the Austrian Science Funds  (FWF  {Grant DOI 10.55776/}I5868), and Grant-in-Aids for Scientific Research (JSPS KAKENHI) Grants No. JP23H03817, and No. JP24K17014. The authors notice that their international collaboration on this specific topic started before
February 2022, hence not violating the current regulatory prescriptions. The calculations have been done in part on the Vienna
Scientific Cluster (VSC).

For the purpose of open access, the authors have applied a CC BY public copyright license to any Author Accepted Manuscript version arising from this submission.



\bibliography{main,addref_metallic-PAM}
\end{document}


\title{Supplementary material for ``Two distinct quantum critical behaviors in the doped two-dimensional periodic Anderson model"}

\author{M. Kitatani$^{a}$, T. Sch\"afer$^{b}$, A. A. Katanin$^{c,d}$, A. Toschi$^e$ and K. Held$^e$}

\affiliation{{$^a$}Department of Material Science, University of Hyogo, Ako, Hyogo 678-1297, Japan}
\affiliation{$^b$Max Planck Institute for Solid State Research, Heisenbergstra{\ss}e 1, 70569 Stuttgart, Germany}
\affiliation{$^c$Center for Photonics and 2D Materials, Moscow Institute of Physics and Technology, Institutsky lane 9, Dolgoprudny, 141700, Moscow region, Russia}
\affiliation{$^d$ Institute of Metal Physics, 620990, Kovalevskaya str. 18, Ekaterinburg, Russia}
\affiliation{$^e$Institute of Solid State Physics, TU Wien, 1040 Vienna, Austria}

\date{\today}

\begin{abstract}
In Section S. 1, we present how the correlation length is extracted from the D$\Gamma$A susceptibilities. In Section S. 2, we show the D$\Gamma$A susceptibilities for different $V$ than those shown in the main text.
\end{abstract}

\maketitle

\section{Extraction of the correlation length $\xi$}
\begin{figure}[htbp]
        \includegraphics[width=\linewidth,angle=0]{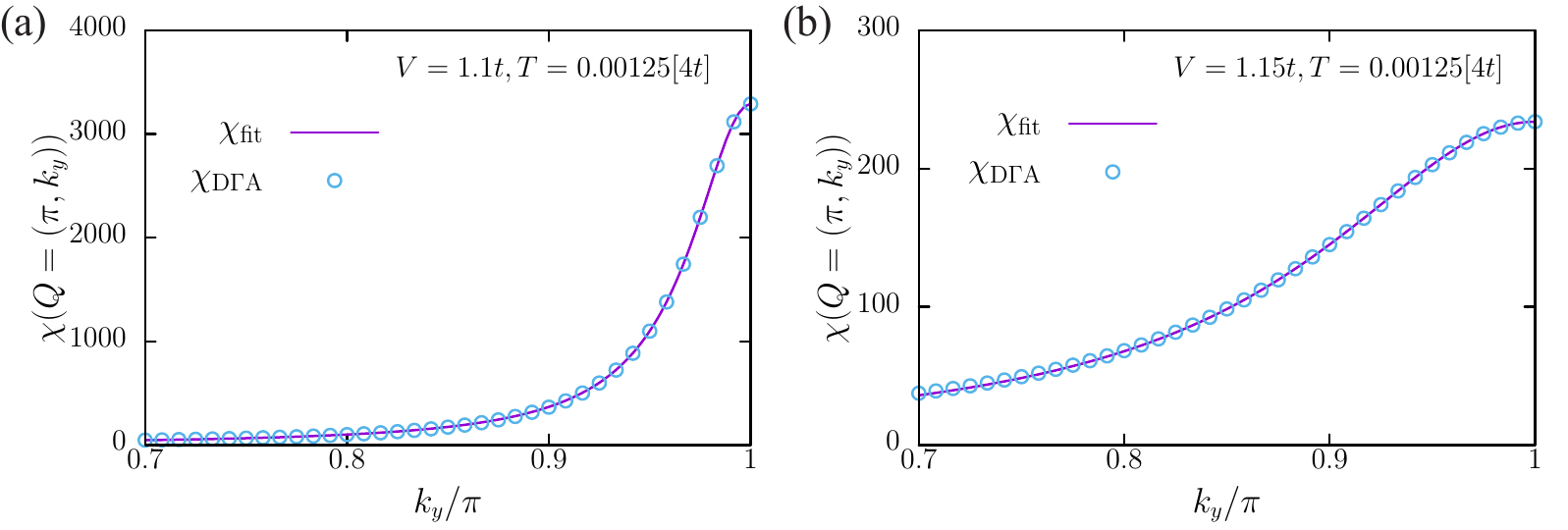}
        \caption{Comparison of the spin susceptibility between the D$\Gamma$A result $\chi_{{\rm D}\Gamma{\rm A}}$ and $\chi_{\rm fit}$ for $T=0.00125[4t]$, (a) $V=1.1t$ and (b) $V=1.15t$.}
        \label{fig:chifit}
\end{figure}

To extract the correlation length $\xi$ from the D$\Gamma$A spin susceptibility, we fit the formula $\chi_{\rm fit}=\frac{A}{(\mathbf{Q}-\mathbf{k})^2+\xi^{-2}}$ to the $\chi_{{\rm D}\Gamma{\rm A}}(\pi,k_y)$ susceptibility using five data points near the peak position $\mathbf{Q}=(\pi,\pi)$. Figure~\ref{fig:chifit} shows an example of the fitting. Please note that we obtained the commensurate fluctuation $\mathbf{Q}=(\pi,\pi)$ for systems with $V \gtrsim V_{\rm KBD}$, on which we performed this fitting. 

\section{D$\Gamma$A spin susceptibilities for different hybridization $V$}
\begin{figure}[htbp]
        \includegraphics[width=\linewidth,angle=0]{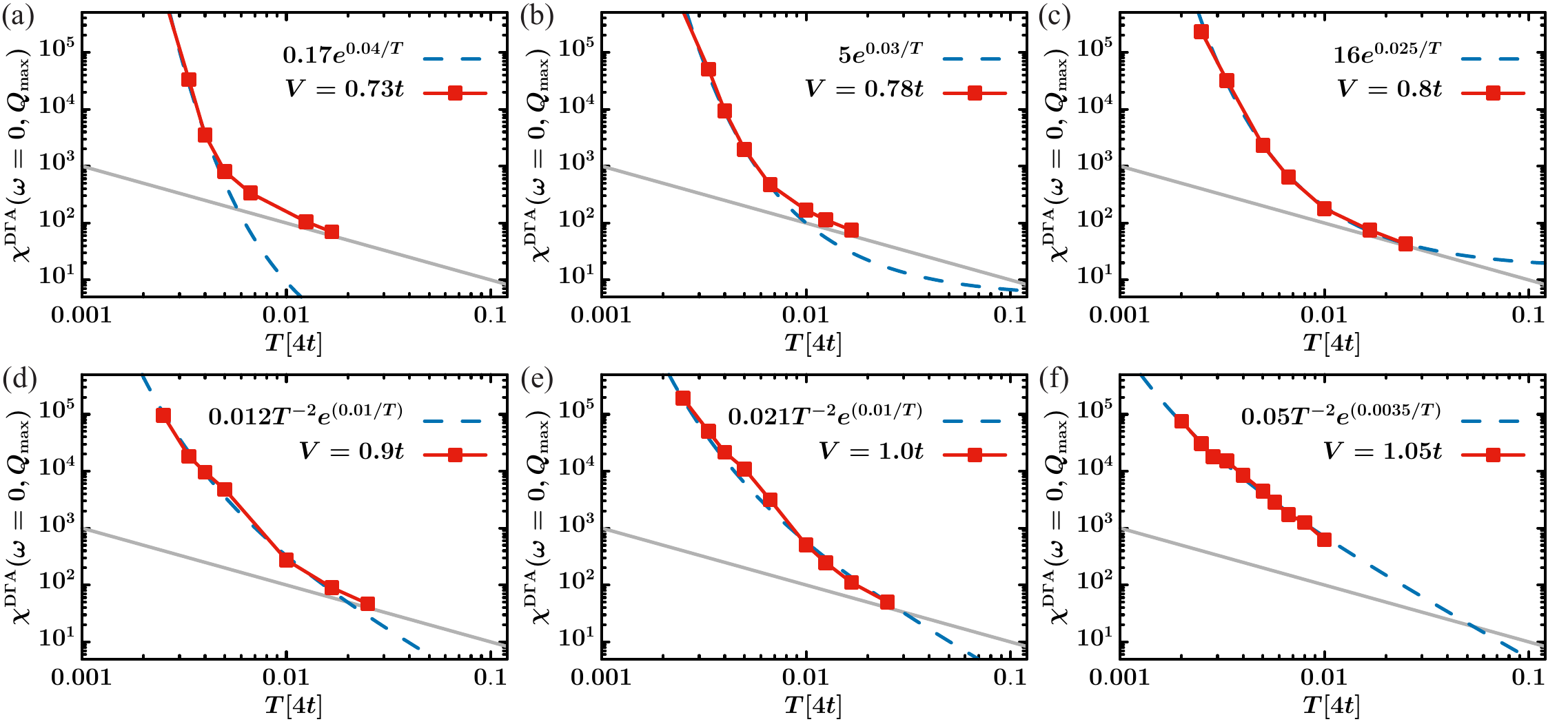}
        \caption{Spin susceptibilities (on a log-log scale) in D$\Gamma$A for (a) $V=0.73t$, (b) $V=0.78t$, (c) $V=0.8t$, (d) $V=0.9t$, (e) $V=1.0t$, and (f) $V=1.05t$.}
        \label{fig:chiTfit}
\end{figure}

In this section, we present additional D$\Gamma$A spin susceptibility results. Fig.~\ref{fig:chiTfit} shows the temperature dependence of $\chi(\mathbf{Q}_{\rm max})$ for $V=0.73t, 0.78t, 0.8t, 0.9t, 1.0t, 1.05t$ (same figures as Fig.~3 in the main text, but here, for different $V$s). As discussed in the main text, we fit the exponential behavior with $\chi \propto e^{\Delta/T}$ for $V=0.73t,0.78t,0.8t$ and $\chi \propto T^{-2} e^{\Delta/T}$ for $V=0.9t,1.0t,1.05t$. The fitted exponents $\Delta$ are shown in Fig.~4 in the main text. 
